# Deep learning enables automated assessments of inflammatory response in zebrafish exposed to different pollutants

Lulu Xu[1], Peiwu Qin[1,2,3,4*], Zhenglin Chen[1,2,3,4*], Jiaqi Yang[5*]

[1]Key Laboratory of Imaging Diagnosis and Minimally Invasive Intervention Research, The Fifth Affiliated Hospital of Wenzhou Medical University, Lishui, Zhejiang Province, 323000, China.

[2]Tsinghua-Berkeley Shenzhen Institute, Tsinghua Shenzhen International Graduate School, Tsinghua University, Shenzhen 518000, Guangdong, China

[3]Institute of Biopharmaceutical and Health Engineering, Shenzhen International Graduate School, Tsinghua University, Shenzhen 518000, Guangdong, China

[4]Key Lab for Industrial Biocatalysis, Ministry of Education, Department of Chemical Engineering, Tsinghua University, Beijing 100084, China

[5]School of Civil & Environmental Engineering, Harbin Institute of Technology (Shenzhen), Shenzhen 518055, China;

* Corresponding author's emails: tongguodong@aa.seu.edu.cn, zhenglin.chen@sz.tsinghua.edu.cn

## Abstract

In the field of environmental toxicology, rapid and precise assessment of the inflammatory response to pollutants in biological models is critical. This study leverages the power of deep learning to enable automated assessments of zebrafish, a model organism widely used for its translational relevance to human disease pathways. We present an innovative approach to assessing inflammatory responses in zebrafish exposed to various pollutants through an end-to-end deep learning model. The model employs a Unet-based architecture to automatically process high-throughput lateral zebrafish images, segmenting specific regions and quantifying neutrophils as inflammation markers. Alongside imaging, qPCR analysis offers gene expression insights, revealing the molecular impact of exposure on inflammatory pathways.

Moreover, the deep learning model was packaged as a user-friendly executable file (.exe), facilitating widespread application by enabling use on virtually any computer without the need for specialized software or training.

# 1. Introduction

The field of environmental toxicology seeks to comprehend the effects of chemical contaminants on biological organisms and ecosystems. Within this field, the zebrafish (Danio rerio) has emerged as a vertebrate model organism of significant interest within toxicological and pharmacological research area due to its genetic similarity to humans and its transparent embryos [1]. The species' physiological response to various environmental stressors, particularly the inflammatory response to pollutants, provides invaluable insights into human and ecological impacts. Inflammation is a fundamental biological process, representing both a ubiquitous marker of pathogen defense and a harbinger of tissue injury across a broad spectrum of disease states. When the body encounters harmful stimuli, such as invading pathogens, wounding, or damaged cells, the immune system will be activated and an inflammatory response is triggered [2]. Rapid and accurate quantification of inflammatory responses is, therefore, a cornerstone of toxicological research.

However, the traditional methodology of assessing inflammation in zebrafish, involving manual microscopic inspection and scoring, is time-consuming and susceptible to inter-observer variability. It also poses a challenge in handling large-scale studies due to the manual effort involved in quantifying neutrophils—a primary biomarker of inflammation [3]. Neutrophils, polymorphonuclear leukocytes, are the first responders to injury and infection in vertebrates, including zebrafish, making them a significant focus of study in inflammatory responses to environmental pollutants [4].

Recent advancements in deep learning have provided unprecedented opportunities for the automation of image analysis [5], particularly in biological imaging [6]. Convolutional neural networks (CNNs), a class of deep learning algorithms, are adept at handling image data and have achieved state-of-the-art performance in tasks such as image classification, localization, and segmentation [7]. The Unet, a CNN variant known for its efficacy in medical image segmentation [8], demonstrates a structure specifically designed to excel in tasks where the output is a high-resolution segmentation map.

In this study, we aim to mitigate the aforementioned challenges by implementing a Unet-based deep learning model for the automated assessment of inflammatory responses in zebrafish. The deep learning framework is trained to accurately perform segmentation of anatomical regions of interest in the zebrafish larvae and to quantify neutrophil infiltration—serving as a surrogate measurement of the inflammatory response induced by exposure to a variety of environmental pollutants.

Given the complex nature of pollutants and their potential combinatorial effects, we centered our analysis on nine pollutants, chosen for their varying chemical properties and known or suspected mechanisms of physiological disruption [9]. Among these contaminants are both ubiquitously present pollutants, such as perfluorooctanesulfonate (PFOS), and those predominantly found in specific regions or applications, such as atrazine, a prevalent agricultural herbicide [10].

To validate our image-based assessment of inflammation [11], we parallel our study with a qPCR analysis [12]. In this molecular approach, we quantify the transcript levels of several inflammatory cytokines and biomarkers, including interleukin-1 beta (IL-1β), interleukin-6 (IL-6), interleukin-10 (IL-10), tumor necrosis factor-alpha (TNF-α), and cyclooxygenase-2 (COX-2) [13]. The expression of these genes serves as a biochemical gauge of the inflammatory state of the organism and further provides insights into the physiological mechanisms underpinning the responses to toxin exposure [14].

A critical aspect of our work is the transformation of the deep learning model into a standalone executable file, thus circumventing the conventional need for deployment on a high-powered workstation with specialized software requirements. By converting the model into a .exe format executable under the Microsoft Windows operating system, we drastically widen the utility of this advanced image analysis technique, bringing it into regular desktop environments of researchers [15].

Our integrative approach, employing both state-of-the-art image processing and molecular validation via qPCR, presents an innovative methodology in environmental toxicology. It paves the way toward scalable, reproducible, and accurate assessments of the impact of environmental pollutants on living organisms, enabling faster and more comprehensive evaluations.

## 2. Materials and Methods

### 2.1 Zebrafish Maintenance and Exposure

The zebrafish (Danio rerio) is an established aquatic model organism in the field of toxicological research, and proper husbandry is essential for obtaining reliable results [16]. Zebrafish colonies were maintained in a recirculating aquaculture system with parameters strictly controlled for temperature (28±1°C), pH (7.0–7.5), and conductivity (500–800 μS/cm) to ensure optimal health and reproductive conditions [17]. The system was on a 14:10 hour light-dark cycle, consistent with zebrafish circadian biology to promote regular behavioral and physiological cycles [18].

Zebrafish aged two days post-fertilization (dpf2) were selected for these experiments due to their developmental stage which allows for the evaluation of both morphological and initial physiological responses to irritants [19].

The pollutants were formulated into an E3 mixture containing 5mM NaCl, 0.17mM KCl, 0.33mM CaCl2, and 0.33mM MgSO4 with a 0.1% v/v concentration of dimethyl sulfoxide (DMSO) as the carrier to ensure solubility while maintaining DMSO levels sufficiently low to eliminate concerns regarding solvent toxicity [20]. Gradient dilution for all samples was performed using this E3 solution [21].

The nine selected pollutants encompassed a range of typical environmental toxicants, such as heavy metals, plasticizers, pesticides, and industrial byproducts [22]. Concentrations chosen for exposures were informed by both the literature and preliminary range-finding tests to ascertain sub-lethal levels for zebrafish embryos [23]. Specifically, these levels were gauged to elicit a biological response without causing acute lethality or severe developmental defects, hence falling within the range of probable environmental exposures [24].

The nine pollutants tested, along with their corresponding gradient concentrations, are meticulously detailed in Table 1 below [25]. Each pollutant concentration was chosen to reflect relevant environmental exposures, ranging from sublethal to potentially lethal doses, to cover a broad spectrum of possible effects [26].

Table 1. Pollutants and Corresponding Gradient Concentrations Used for Zebrafish Exposure.

| Type | Full name | Abbreviation | Concentration |
|---|---|---|---|
| Endocrine Disruptors | Perfluorooctanesulfonate | PFOS | 1,2,5,10,15 μM |
| | Short-chain Chlorinated Paraffins | SCCPs | 1,5,10,15,20 mg/L |
| | Atrazine | ATZ | 10,100,200,500,1000 ug/L |
| Brominated Flame Retardants | Hexabromocyclododecanes | HBCD | 0.01,0.1,0.5,1.25,2.5 μM |
| | Bisphenol A | BPA | 5,10,15,20,25 μM |
| | Decabromodiphenyl Ether | BDE-209 | 0.1,1,2.5,5 μM |
| Heavy Metal | Mercury Chloride | $HgCl_2$ | 0.03,0.05,0.09,0.15,0.25 mg/L |
| | Cadmium Chloride | $CdCl_2$ | 1,10,50,100,250 μM |
| | Lead Iodide | $PbI_2$ | 0.2,2,5,10,20 μM |

For the exposure period, dpf2 zebrafish were incubated in the pollutant solutions for 24 hours. The aftermath of exposure to the pollutants was investigated under high-resolution microscopy. All procedures followed the guidelines established by the Institutional Animal Care and Use Committee (IACUC), and principles of the 3Rs (Replacement, Reduction, and Refinement) were strictly adhered to minimize animal use and distress [27].

## 2.2 Image Acquisition and Preprocessing

### A. Acquisition of Zebrafish Fluorescence Images

The retrieval of fluorescence images commenced with the utilization of a Nikon stereomicroscope to capture lateral views of the zebrafish specimens [28]. The stereoscope is a sophisticated optical instrument that provides three-dimensional visualization using two separate optical paths [29], delivering depth cues vital for the examination of biological samples [30]. The images were obtained using a high-sensitivity digital camera specifically calibrated for low-light conditions to accurately record the emitted fluorescence signals from the zebrafish [31].

### B. Preprocessing to Enhance Image Quality

Initially, the acquired images suffer from a diminished brightness, rendering them ineffective for direct analysis through deep learning models and also complicating the process of manual annotation necessary for precise region identification [32].

The preprocessing stage involves the transition from the standard RGB color space to the HSV color space, as the latter is considered profoundly effective when performing enhancements contingent on luminance and chromaticity [33]. RGB space defines color in terms of red, green, and blue components, which could be computationally restrictive for certain image processing applications, particularly when adjustments related to brightness or saturation are required [34]. Conversely, HSV simplifies these manipulations by separating image intensity (Value) from color information (Hue and Saturation) [35].

After the conversion to HSV space, we performed tailored alterations to the Value channel to surmount the issue of underexposure. This adjustment aims to augment the pixel intensity distribution without disrupting the inherent color composition of the image [36].

Simultaneously, adjustments to the Saturation channel are executed to heighten the contrast between vividly and dimly colored regions, thus further refining the image's suitability for subsequent analytical procedures. Saturation control selectively amplifies colorfulness, which is imperative for distinguishing anatomical regions of zebrafish exhibiting minute differential color signatures [37].

To illustrate the effects of the image processing adjustments made, we present Figure 1, which showcases the comparative results. Through Figure 1, we can discern the practical benefits gained from the image processing steps.

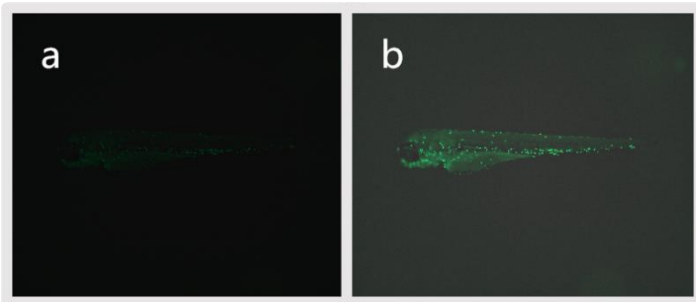

Figure 1: Comparative Images Before and After Processing

After the process, images manifest enhanced luminance and distinction, rendering them apt for deep learning model ingestion and simpler manual annotations [38]. The enforcement of these preprocessing steps safeguards against variability in image acquisition settings, and upsurges overall algorithmic precision and robustness.

## 2.3 Deep Learning Model Design

To detail the workflow and outline the structure of the implemented model, a comprehensive flow chart has been developed. Figure 2 below visually demonstrates the overarching process from image acquisition to data analysis and interpretation, delineating each stage within the operational pipeline of the model [39].

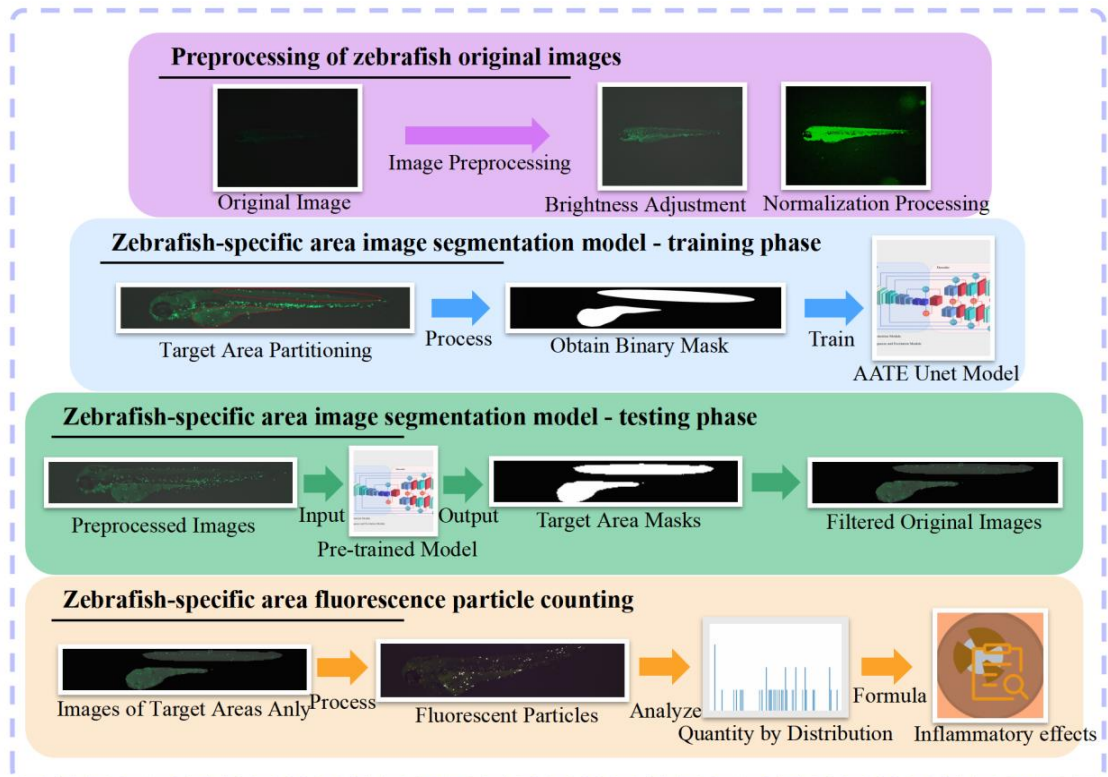

Figure 2: Overall Flow Chart of the Deep Learning Model

Figure 2 serves as an instructive guide, charting the linear progression of tasks aligned with the model's use—beginning with the initial input of biological images, proceeding through the stages of preprocessing and analysis, and culminating with quantification and data synthesis.

### A. Mixed Unet Architecture

Our research pivots around using a novel convolutional neural network structure named the Mixed Unet. This network thrives upon the architecture of the Unet model [40]. The Mixed Unet introduces an additional decoding branch in the expansive path, thereby transmuting the single-decoder paradigm of the classical Unet to a dual-decoder configuration. The synergistic operation of these dual decoders equips the Mixed Unet with a profusion of feature representations, heightening its competence to discern and segment intricate image details [41].

### B. Dataset and Ground Truth Annotation

The neural network was indoctrinated utilizing a dataset of zebrafish images. This set of images was amassed to assist in the assessment of anti-inflammatory properties of cosmetic products and was graciously furnished by Guangzhou Baiyun Meiwan Co., Ltd. Precise quantification of neutrophils within each image was meticulously conducted through manual counting procedures, as neutrophils serve as a key biological marker for identifying inflammatory responses [42].

### C. Region-Specific Focus and Annotation

To explore the inflammatory effect induced by disparate pollutants, it is essential to enumerate neutrophils within particularized biotic regions, mainly the yolk and spine areas of the zebrafish. Due to the innate propensity of neutrophils to migrate towards cutaneous layers in states of inflammation [43], these areas are chosen because of their adjacency to the epidermal surface. Regions such as blood vessels are eschewed, given the high density and consequent enumerative difficulty of neutrophils therein.

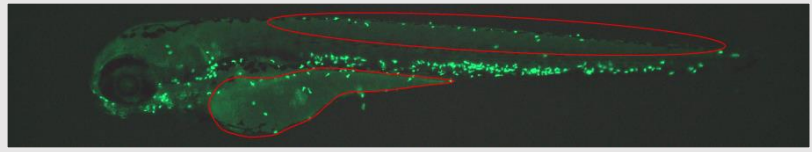

Figure 3: Labeled Yolk and Spine Areas of the Zebrafish

Enumerating neutrophils in these particular biotic regions provides insight into the local inflammatory responses that are trademark reactions to toxic stress. Such targeted analysis necessitates a clearly labeled visual representation of the zebrafish anatomy to facilitate accurate and repeatable identification and counting of neutrophils. As a result, Figure 3 provides a distinct labeling of the yolk and spine areas [44].

### D. Model Training and Objective

Before the initiation of model training, an imperative step is to manually annotate the images demarcating the yolk and spine areas [45]. These critical annotations are infused into the model as labels, facilitating the model's learning towards accurate region-specific segmentation [46]. The training seeks to enable the model to auto-segment the yolk and spine regions, thereby catalyzing the counting of neutrophils in these regions, a process paramount for analyzing the inflammatory reactions of zebrafish [47].

### E. Preliminary Data Annotation Process

In our specific use case, the dataset images of zebrafish were rendered with preliminary annotations delineating the yolk and spinal regions. We have engineers encircling targeted biological areas with a distinguishable red linear outline, shown in Figure 3. This initial manual annotation is the bedrock upon which binary mask inputs for the deep learning model are generated.

### F. Transformation from Linear Outline to Binary Mask

For the Mixed Unet model to correctly interpret the regions of interest from the annotated zebrafish images, a conversion process must be implemented to transpose the red linear outlines into binary masks. This transformation process follows a multi-step methodology:

Step 1: Contour Creation

Utilizing a color thresholding technique, the original images undergo filtration to isolate the red-outlined portions. The red channel is exclusively targeted, enabling the extraction of the contour. However, after this step, non-contiguous dark red artifacts may persist due to variations and imperfections in the red tone of the contour lines employed in the manual annotation process.

Step 2: Residual Cleanup

Cleaning is imperative to refine the extracted contour by eradicating any residual dark red fragments that could adulterate the purity of the binary masks. This is realized through image processing techniques that discern and remove these residual artifacts, ensuring that the non-contour regions of the image are transformed to maximal blackness, thereby establishing a stark contrast with the contour.

Step 3: Solid Filling of Contour

Upon obtaining a clean contour, a filling operation is employed. This task is executed using the `cv2.fillPoly` function from the OpenCV library, a well-regarded open-source computer vision and machine learning software library, to scan for and fill the contour, thereby creating a solid white mask corresponding to the regions that were initially annotated with the red outline.

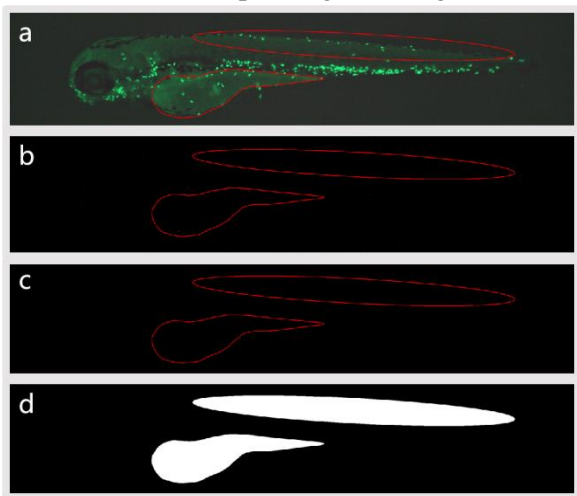

Figure 4: Transformation Steps from Red Outlines to Binary Masks

**G. Model Training and Neutrophil Detection**

With the creation of binary masks complete, these masks coupled with their corresponding original images are utilized as paired training inputs for the Mixed Unet model. The trained model exhibits remarkable segmentation accuracy, cleanly separating the yolk and spine regions from the rest of the image contents.

The segmented output next will be processed to get the quantification of neutrophils, which are identified by their high pixel intensity. Utilizing the SimpleITK toolkit [48], a pixel intensity threshold is set to discern neutrophils as highly bright pixel clusters. Connected clusters that exceed a specified size threshold are deemed as neutrophil entities.

The deployment of this automated neutrophil detection mechanism is then extended to analyze the inflammatory response within zebrafish subject to varying pollutants, facilitating the automatic assessment of environmental toxicology [49].

The design intricacies of the employed deep learning model [50], powered by the tailored preprocessing steps and post-segmentation neutrophil quantification algorithms, allow for a nuanced analysis of inflammatory effects. This paradigm underpins a vital intersection of computational biology and environmental science, where the automated, high-throughput analysis could yield significant insights.

## 2.4 qPCR Analysis

In this study, quantitative Polymerase Chain Reaction (qPCR) is utilized as a pivotal molecular technique for quantifying the expression levels of certain cytokines and inflammatory mediators within biological samples. The cytokines analyzed include interleukin-1 beta (IL-1β), interleukin-6 (IL-6), interleukin-10 (IL-10), tumor necrosis factor-alpha (TNF-α), and Cyclooxygenase-2 (COX-2). The specific emphasis on these cytokines stems from their critical roles as mediators and regulators in inflammatory responses [51]. The primer sequences for qPCR are shown in Table 2.

Table 2: Primer Sequences for qPCR

| Gene | Direction | Sequence |
|---|---|---|
| IL-1β | F | TCGCCCAGTGCTCCGGCTAC |
|  | R | GCAGCTGGTCGTATCCGTTTGG |
| IL-6 | F | TCAACTTCTCCAGCGTGATG |
|  | R | TCTTTCCCTCTTTTCCTCCTG |
| IL-10 | F | TCACGTCATGAACGAGATCC |
|  | R | CCTCTTGCATTTCACCATATCC |
| TNF-α | F | AGGAACAAGTGCTTATGAGCCATGC |
|  | R | AAATGGAAGGCAGCGCCGAG |
| COX-2 | F | AACTAGGATTCCAAGACGCAGCATC |
|  | R | AAATAAGAATGATGGCCGGAAGG |

By exploring the expression profiles of these inflammatory genes, this study aims to clarify the modulation of inflammatory pathways in zebrafish and to uncover the biological impacts of prospective anti-inflammatory agents [52].

### A. Primer and Probe Design

The accuracy of qPCR is significantly dependent on the design of primers and probes, which are short strands of nucleic acids that recognize and bind to their complementary sequences on the DNA template [53]. These primers flank the DNA sequence of interest, thus delineating the region for amplification [54]. Additionally, probes labeled with fluorescent reporters are employed to emit fluorescence upon hybridization to the target, allowing real-time detection of the amplification products [55].

### B. Amplification and Quantification Protocols

The qPCR process undergoes a series of thermal cycling steps involving denaturation, annealing, and extension phases. Upon denaturation, the doubly helical structure of the template DNA is melted to yield single-stranded molecules. Subsequent annealing allows the primers and probes to hybridize to these single strands at lower temperatures. The extension phase, typically at a slightly higher temperature, is where the DNA polymerase enzyme synthesizes new strands of DNA, complementary to the original template [56]. The amount of fluorescence emitted correlates directly with the amount of DNA being amplified [57].

### C. Analysis of Inflammatory Cytokines

The qPCR process provides an efficient and sensitive means to quantify the transcriptional levels of specific cytokines, which are indicative of inflammation when regulated aberrantly [58]. The amplification of IL-1β, IL-6, IL-10, TNF-α, and COX-2 mRNA is particularly essential due to their substantial relevance in the signal transduction pathways responsible for the development and resolution of inflammation. IL-1β and TNF-α are primary pro-inflammatory cytokines, orchestrating acute phase responses, whereas IL-6 has wide-ranging effects, both pro- and anti-inflammatory [59]. IL-10, conversely, is an anti-inflammatory cytokine that modulates the intensity of the immune response [60]. COX-2 is an inducible enzyme catalyzing the production of prostanoids, mediating various physiological processes including inflammation [61].

### D. Data Interpretation and Significance

The difference in cycle threshold (CT) values between the target and reference genes allows the determination of relative gene expression levels through the $2^{-\Delta\Delta CT}$ method [62]. The relative quantification of the cytokines mentioned provides insights into the underlying inflammatory processes induced by various stimuli or pathological conditions. This quantitative measure equips researchers with the ability to evaluate the modulation of the inflammatory response within the studied biological context.

## 2.5 Software Deployment

The zebrafish image analysis software is designed for the identification and quantification of the neutrophils within specific areas of zebrafish. In the user interface, the 'Select Picture' button enables users to navigate through their local storage to select multiple image files at once for analysis. The 'Next Picture' button is designed to trigger the display of individual analysis results sequentially.

The user interface (UI) of the software, which is depicted in Figure 5, provides an intuitive layout for users to interact with the program [63].

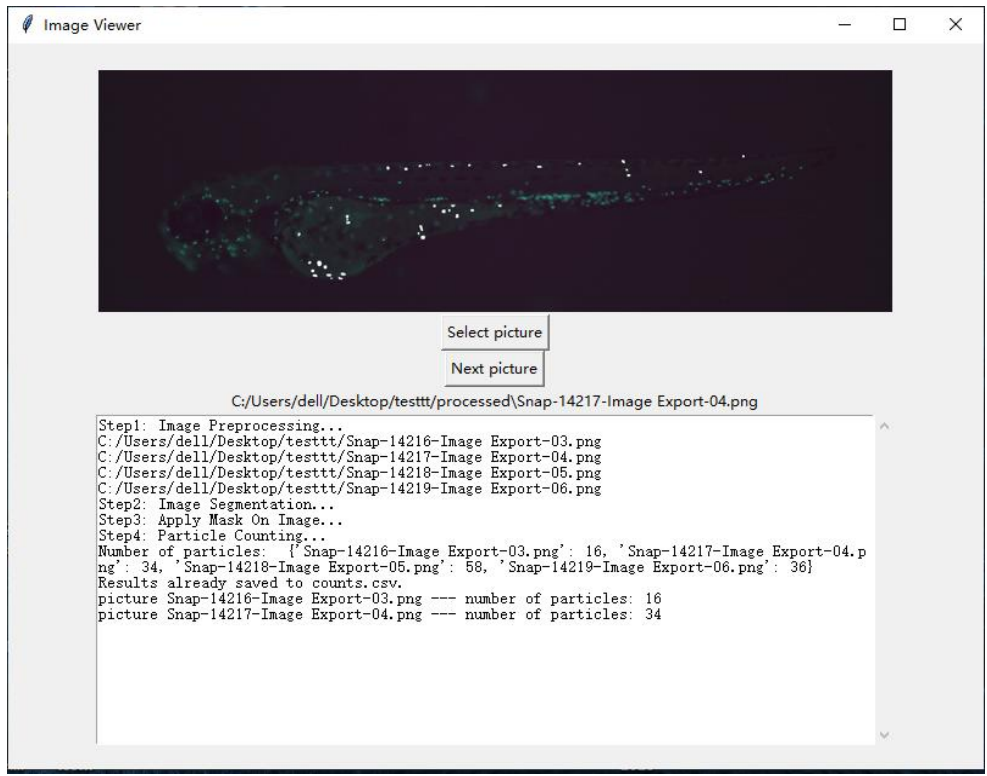

Figure 5: Software Interface for Neutrophil Quantification

Upon the selection of an image, proprietary algorithms are deployed to recognize and mark the position of neutrophils in the original image. The processed image concurrently gets displayed in the software's interface and is automatically stored in a predetermined 'Processed' directory for user convenience and later access [64].

Following the processing, quantifying neutrophils becomes a subsequent requisite action [65]. A counting algorithm, which is a part of the deployed image processing suite, enumerates the detected cells [66]. Another essential feature of the software is the generation of an Excel file. This Excel file serves as a summary for collating the neutrophil counts from all analyzed images.

# 3. Results

Our investigations apply computational techniques to the study of zebrafish in biological and environmental research. We elaborate on our findings concerning the efficacy of the proposed methodologies employed in the segmentation and analysis of zebrafish fluorescence images, the precision of an automated neutrophil counting algorithm, and the effects of various pollutant concentrations on zebrafish inflammatory responses, both at a cellular level and in terms of gene expression.

**A. Segmentation Efficacy of the Mixed Unet Model**

We examine the effectiveness of a Mixed Unet model, a modified version of the conventional U-Net model augmented to enhance feature extraction through hybridized convolutional layers that capture diversified spatial and contextual detail within biological image data. The preliminary results from the Mixed Unet model for the segmentation of zebrafish fluorescence images, specifically focusing on the yolk sac and spine regions, indicate a high congruence with manually annotated benchmarks. The discrepancy between the automated segmentation by the Mixed Unet and the expert manual annotations was minimally divergent, validating the former's precision in biological imaging analysis.

The preliminary results from the Mixed Unet model are presented in Figure 6, which displays an array of segmentation outputs to illustrate the model's efficacy. The images in this Figure are neatly arranged into three columns for comparative purposes. The left column featured the preprocessed fluorescent images. The middle column showcased the ground truth segments, manually annotated by expert engineers to serve as a benchmark for accuracy. The right column revealed the segmentation results obtained from the Mixed Unet model.

A visual inspection of Figure 5 elucidates that the model's segmentation results bear a remarkable resemblance to the ground truth. The disparities between the two are minor. This near-perfect match between manual and automated outputs accentuates the precision with which the Mixed Unet model operates, reinforcing its applicability as a tool for high-fidelity biological imaging analysis.

To quantify the performance and validate the visual findings, we computed the Mean Intersection-over-Union (MIoU) metric across the entire dataset. This metric, a staple in assessing the accuracy of image segmentation tasks, compares the pixel-wise overlap between the predicted segmentation and the ground truth. An MIoU score of 1 would indicate perfect congruence with no divergence at all. In our extensive dataset, the Mixed Unet model achieved an exceptional MIoU score of 0.9865, demonstrating that the model's automated segmentations closely mirrored those of the manual annotations. This high level of accuracy validates the model's utility.

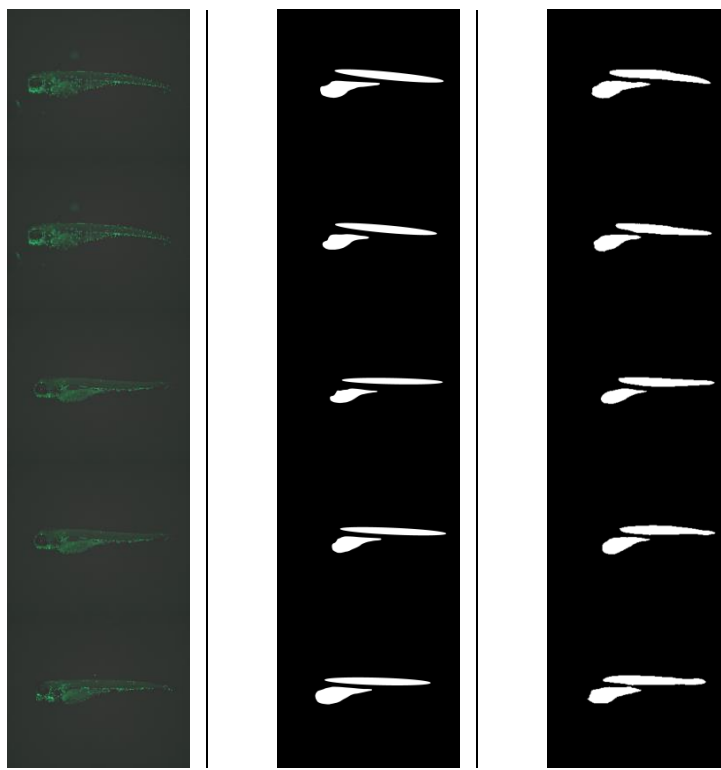

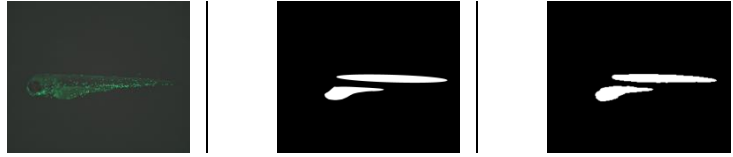
Figure 6: The preliminary results from the Mixed Unet model

**B. Neutrophil Counting Algorithm Efficacy**

We assessed our algorithm devised for enumerating neutrophils within the complex milieu of zebrafish fluorescence images. The algorithm's predictive capacity was critically compared to traditional manual counting techniques. It was found that the probability of achieving an error margin within 10% when compared to manual counting methods stood at 75%.

Considering the inherent inaccuracy in human annotation, this error range is deemed highly precise and acceptable. The likelihood of the prediction falling within the 10% error margin was substantially high, reflecting a proficient algorithmic detection consistent with the ambiguity levels of manual counting procedures.

The visual confirmation of the algorithm's accuracy is provided in Figure 7, which serves as a comparison highlighting the recognition capabilities of our software. In the figure, the right side features five images that present the results of the automatic neutrophil counting carried out by the software. Each image vividly marks the identified neutrophils, denoted by the highlighted regions superimposed onto the original image. On the left side of Figure 7, the original, unannotated fluorescence images are displayed, allowing for a direct visual comparison between the raw data and the processed results.

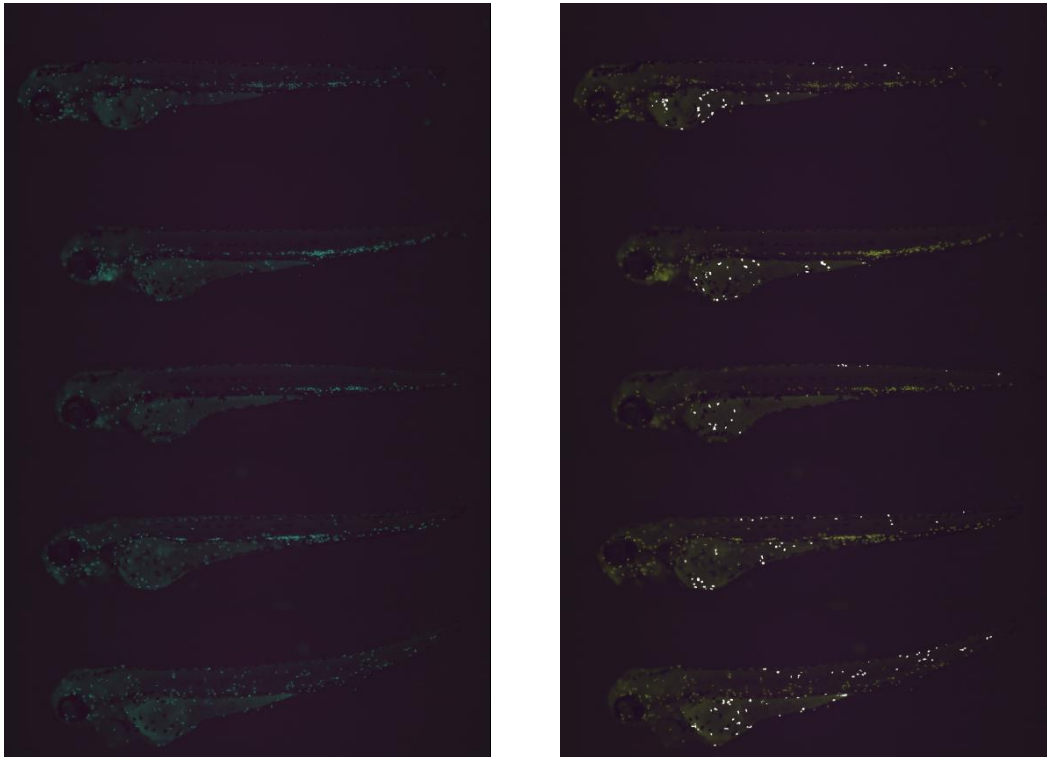
Figure 7: Visual comparison between the raw data and the processed results

### C. Inflammatory Marker Expression Via qPCR Analysis

Our multifaceted study on the impact of pollutants on zebrafish not only involved cellular-level analysis through automated neutrophil counts but also extended to the molecular level through the deployment of quantitative PCR (qPCR) techniques [67]. The latter aimed to elucidate the correlative relationship between the different concentrations of the indicated pollutants and the expression levels of specific inflammatory marker genes within the zebrafish.

In the pursuit of comprehensive insights, we utilized qPCR to amplify and quantify cDNA, which was reverse-transcribed from the zebrafish mRNA [68]. This in-depth approach provided a nuanced view of the genetic response to each pollutant across a spectrum of dosages. By carefully measuring the relative increase in gene expression consequent to the exposure [69], the results exhibited a concentration-dependent induction of these critical inflammatory markers [70]. Such molecular findings complemented the physical observations gained from the neutrophil counts, thus enabling a more integrative understanding of the biological implications of pollutant exposure [71].

The qPCR experiments were meticulously carried out to quantify the upregulation of inflammatory markers in zebrafish in response to the pollutants, with data gathered across three distinct concentration gradients. The ensuing analysis delineated a pronounced concentration-dependent relationship. This was particularly evident in the direct association noted between heightened concentrations of pollutants and the consequential stimulation of inflammatory genes. The clear correlation that surfaced through our qPCR assays bridged the gap between the organism-level data, represented by the neutrophil response, and the underlying molecular changes, offering a compelling molecular perspective that correlated directly with the observed physiological responses.

By adopting this twofold approach, our research provided a multi-angled view of the effects pollutants exert on zebrafish, addressing both the macroscopic manifestations and the subtle molecular shifts. This dual analysis, with the meticulous counting of neutrophils and the sensitive quantification of gene expression, crystallized the correlation between pollutant dosage and biological response. Our findings contribute significantly to the assessment of environmental risk, establishing a robust methodological basis for assessing aquatic health and safety in the context of pollution exposure and reinforcing the importance of integrated studies in environmental toxicology.

Table 3 presents experimental data encompassing neutrophil granulocyte counts and fold changes in the expression of five different genes: IL-1β, IL-6, IL-10, TNF-α, and COX-2. These genes play pivotal roles in inflammatory responses, while neutrophils are key cells involved in such reactions.

Table 3 presents a comparative analysis of the neutrophil granulocyte counts obtained from the zebrafish-specific software and the fold changes in gene expression (IL-1β, IL-6, IL-10, TNF-α, and COX-2) for pollutant-concentration combinations. The software-derived neutrophil counts represent the number of neutrophil granulocytes in a specific region of zebrafish exposed to the corresponding pollutants and concentrations, while the fold changes in gene expression are derived from qPCR experiments. The software's predictions were generated using the same pollutant exposure conditions and concentrations as those in the qPCR experiments.

Each row in Table 3 represents a different concentration level, with concentration 1 representing the lowest concentration in the concentration gradient and concentration 5 representing the highest concentration. The "Neutrophil Granulocyte Counts (Software)" column displays the average number of neutrophils counted in the designated region of the zebrafish sample. The "Fold Change in Gene Expression" columns list the mean fold changes in expression levels across the nine pollutants at the same concentration level. The fold changes are calculated relative to a control group (untreated zebrafish group).

Table 3 Comparative analysis of neutrophil granulocyte counts and gene expression data

| Concentration | Neutrophil Granulocyte Counts (Average) | Fold Change in IL-1β Expression (Average) | Fold Change in IL-6 Expression (Average) | Fold Change in IL-10 Expression (Average) | Fold Change in TNF-α Expression (Average) | Fold Change in COX-2 Expression (Average) |
|---|---|---|---|---|---|---|
| Concentration 1 | 24.29 | 1.53 | 1.30 | 1.49 | 1.38 | 0.27 |
| Concentration 2 | 26.81 | 1.65 | 1.48 | 1.96 | 1.74 | 0.33 |
| Concentration 3 | 30.81 | 1.74 | 1.73 | 2.56 | 1.77 | 0.49 |
| Concentration 4 | 36.65 | 1.87 | 1.87 | 2.96 | 2.28 | 0.51 |
| Concentration 5 | 39.12 | 1.97 | 2.09 | 3.50 | 2.48 | 0.62 |

Figure 8 provides a direct comparison between the experimentally measured average neutrophil counts and the corresponding mean gene expression changes for each concentration level across the nine pollutants. The blue area is the number of neutrophil granulocytes, corresponding to the vertical axis on the right. Five different colored lines represent the fold change in the expression of five genes, corresponding to the vertical axis on the left. The horizontal axis represents the concentration gradient.

To analyze the data provided above, we have calculated the Pearson correlation coefficients between neutrophil counts and the fold changes in gene expression, shown in Table 4. The Pearson correlation coefficient is a statistical measure that quantifies the linear relationship between two variables, ranging from -1 to 1. A value of 1 indicates a perfect positive correlation, -1 indicates a perfect negative correlation, and 0 suggests no correlation.

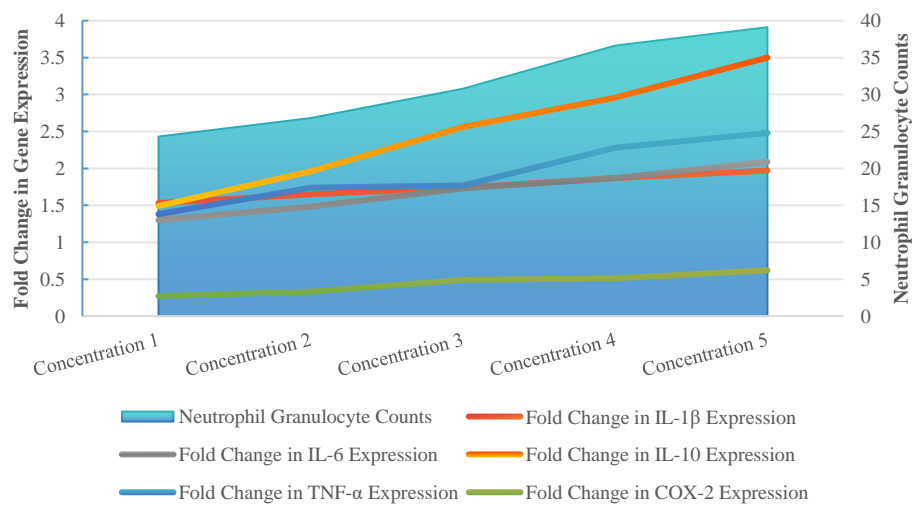

Figure 8: Visual comparison between the neutrophil granulocyte counts data and the fold change in gene expression data.

From Table 4, it is evident that the Pearson correlation coefficients between neutrophil counts and the fold changes in IL-1β, IL-6, IL-10, and TNF-α expression are all close to 1, indicating strong positive correlations [72]. This suggests that as neutrophil counts increase, the expression levels of these genes also increase, consonant with their known roles in inflammatory responses.

Overall, the data in this table support the positive correlation between neutrophil counts and the fold changes in these gene expressions [73], further emphasizing the AATE-UNet model and the automatic software can effectively reflect the expression profiles of the investigated genes by the mean neutrophil granulocyte counts [74].

Table 4 Pearson correlation coefficient of neutrophil granulocyte counts and gene expression data

| Pearson Correlation Coefficient | Fold Change in IL-1β Expression | Fold Change in IL-6 Expression | Fold Change in IL-10 Expression | Fold Change in TNF-α Expression | Fold Change in COX-2 Expression |
| --- | --- | --- | --- | --- | --- |
| Neutrophil Granulocyte Counts | 0.99 | 0.98 | 0.99 | 0.98 | 0.96 |

## 4. Discussion

In the ensuing discussion, we integrate the aforementioned results to assess the implications and the significance of the findings from the Mixed Unet model segmentation performance, neutrophil counting accuracy, and the implications of pollutant concentrations on inflammatory response in zebrafish [75].

The Mixed Unet model has emerged as a potent tool for segmenting complex biological images, as reflected in the segmentation outcomes achieved for the yolk sac and spinal regions in zebrafish [76]. Not only does this underscore the efficacy of the model in detailed anatomical delineation, but also emphasizes the potential for broader application within biological imaging, where precision is critical [77].

The neutrophil counting algorithm's high accuracy, having less than a 10% error margin, underscores the viability of automated processes in replacing time-consuming manual processes [78]. Importantly, it also lays the groundwork for further research into automated biological image analysis algorithms which can operate with minimal error margins akin to manual counting, but with much greater efficiency [79].

The elucidation of the relationship between the nine different pollutants and neutrophil counts presents a stepping stone for a more comprehensive understanding of zebrafish as a model organism for studying inflammation and the toxicological impacts of various substances. For example, the significant response at 15μM PFOS concentration signifies a key benchmark for pollutant impact studies.

Finally, the qPCR results deliver a quantitative basis for understanding how pollutant concentrations affect the expression of inflammation markers in zebrafish. This not only complements the neutrophil count findings but also allows for a more holistic view of the inflammatory response, bridging gaps between cellular responses and molecular biology.

The outcomes of these studies are pivotal for the advancement of high-throughput screening methods in environmental pollution monitoring and biomedical research and raise the prospect for deep-learning-based approaches to become a mainstay in complex biological analysis. Future works should aim at enhancing the models and methods for broader pollutant spectrum analysis, with larger datasets to further validate and possibly improve upon the established results.

## 5. Conclusion

In conclusion, our comprehensive analysis involving advanced deep learning models, automated neutrophil counting algorithms, and systematic qPCR assays has highlighted critical insights into the effects of environmental pollutants on zebrafish. These insights further the understanding of zebrafish inflammatory responses and establish a foundation for future research on environmental toxicity and biological response mechanisms. The automation and quantification of these processes exhibit substantial progress in the intersection of computational biology and toxicological studies, presenting a new era of precise, efficient, and reproducible research methodologies.